\documentclass[amsmath,amssymb]{revtex4-1}

\usepackage{amssymb}
\usepackage{amsthm}
\usepackage{amsmath}
\usepackage{mathrsfs}
\usepackage{subfig}
\usepackage{multirow}
\usepackage{xcolor}
\usepackage{lineno}
\usepackage{ulem}
\usepackage{graphicx}

\newcommand{\myparagraph}[1]{\paragraph{#1}\mbox{}\\}

\begin{document}

\title{Modeling rain-driven overland flow: empirical versus analytical friction terms in the shallow water approximation}

\author{G. Kirstetter$^1$, J. Hu$^2$, O. Delestre$^2$, F. Darboux$^3$, P.-Y. Lagr\'ee$^1$, S. Popinet$^1$, J.-M. Fullana$^1$ and C. Josserand$^1$} 
\affiliation{$1$ Sorbonne Universit\'es, UPMC Univ Paris 06, CNRS, UMR 7190, Institut Jean Le Rond d'Alembert, F-75005 Paris, France\\
$2$ Laboratoire de Math\'ematiques J.A. Dieudonn\'e - Polytech Nice-Sophia , Universit\'e de Nice - Sophia Antipolis, CNRS, UMR 7351, Parc Valrose, 06108 Nice cedex 02, France\\
$3$ Inra, UR 0272, UR Science du sol, Centre de recherche Val de Loire, CS 40001, F-45075 Orl\'eans Cedex 2, France\\
Presently at: Inra, Universit\'e de Lorraine, UMR 1120, Laboratoire Sols et Environnement,F-54505 Vand½uvre-les-Nancy, France}

\begin{abstract}
Modeling and simulating overland flow fed by rainfall is a common issue in watershed surface hydrology. Modelers have to choose among various friction models when defining their simulation framework. The purpose of this work is to compare the simulation quality for the Manning, Darcy-Weisbach, and Poiseuille friction models on the simple case of a constant rain  on a thin experimental flume. Results show that the usual friction law of Manning is not suitable for this type of flow. The Poiseuille friction model gave the best results both on the flux at the outlet and the velocity and depth profile along the flume. The Darcy-Weisbach model shows good results for laminar flow. Additional testing should be carried out for turbulent cases.
\end{abstract}

\maketitle

\section{Introduction}
The rain falling on agricultural fields produces overland flows, which lead to soil erosion (\cite{Moss1979}, \cite{Kinnell1999}), pollutant transport (\cite{Cai2007}, \cite{Benkhaldoun2007}) and flood events downstream (\cite{Cea2010}, \cite{An2015}). 
To prevent and understand these often undesirable effects, rain-induced flows have to be modeled accurately, thanks in particular to numerical simulations. As long as the flows have a horizontal length scale larger than the vertical one, the vertical velocity profile can be integrated, leading to a 2D system of equations, called the shallow-water equations (\cite{DeSaint-Venant1871}). 
Such shallow-water equations are commonly used for modeling overland flow (\textit{e.g.} \cite{Smith2007a}), tsunamis (\textit{e.g.} \cite{Popinet2011}), dam breaks and flood events (\textit{e.g.} \cite{An2015}) or river flooding (\textit{e.g.} \cite{Bates2010}), which are generally flows at high Reynolds numbers. Because numerical simulations of such systems play a significant role in government decision-making to prevent or control inundation risks, it is crucial to properly model the underlying physical mechanisms as well as develop accurate and validated numerical schemes. 

One of the key points in the shallow-water framework is the effective friction term which depends on the assumption made for the vertical velocity profile. This friction term depends on several parameters, but principally on the dynamical characteristics of the flow (\textit{i.e.} laminar or turbulent). In general, because the flows are at high Reynolds numbers and also because of complex topography and scale effects (see for instance \cite{Smith2014}), empirical laws are used, in particular the Darcy-Weisbach and the Manning models (see for instance \cite{TeChow1959}, \cite{Smith2007a}, \cite{Violet}, \cite{Chanson2004} and \cite{An2015}).  However, it is important to notice that for rain-induced flow, the thin  liquid films involved have small Reynolds numbers. Hence, the use of turbulent modeling is questionable, compared to the classical laminar friction term deduced from a Poiseuille  velocity profile. Moreover, quantitative experiments are still rare (\cite{Esteves2000}), underlying the need for
systematic quantitative comparisons between numerical models and experimental measures.

In this paper, we focus on an  ``ideal rain" over a rough impermeable substrate. Experimental laboratory results are compared with numerical results of the shallow-water equations using both empirical (Darcy-Weisbach and Manning models) and a laminar (Poiseuille model) friction terms. We will show that in this case, the laminar version of the shallow-water equations is the suitable model for overland flows that can be generalized using a Darcy-Weisbach approach.. The configuration studied is presented in the next section as well as the experimental setup. The numerical methods are described in section III, as well as validating cases. The numerical results are compared with the experimental measurements in section IV, and a general discussion is then given.

\section{Materials and methods}
\subsection{The ``ideal rain" case}

\begin{figure*}
\center
\includegraphics{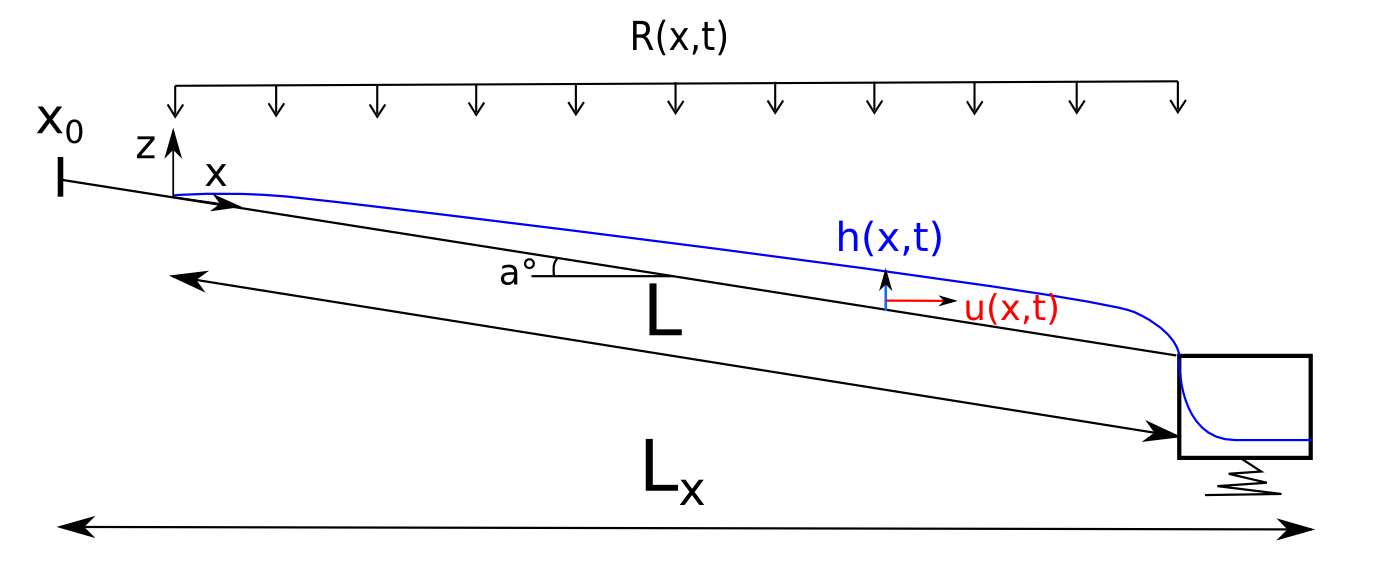}
\caption{The ``ideal rain" case: an homogeneous rain is falling on a tilted flume, producing overland flow.\label{fig_schema}}
\end{figure*}

The numerical simulations of the shallow-water equations are compared with experimental measurements on an ideal configuration of overland flow produced by rain. 
Real cases in nature are complicated to model for various reasons: firstly the topography is often complex and not always well-known; then rainfall is usually not  measured everywhere; finally many different physical mechanisms are imbricated in nature (rain, erosion, infiltration, \textit{etc}). Dedicated experiments where these different effects can be isolated then need to be designed. We focus here on an ideal case of rain falling on a flat impermeable surface as shown in Fig.~\ref{fig_schema}. The same experimental setup was used before to evaluate the validity of numerical schemes in \cite{Delestre2009}. The flat topography is tilted by an angle $a$ and a constant rain intensity equal to $I$ ($mm.h^{-1}$) is imposed. The flume has a length $L=4.04 \ m$ (direction x) and width $l= 11.5 \ cm$ (direction y), and is initially dry. The rain leads to an overland flow which is characterized by $h_{2D}(x,y,t)$ the water depth and $u_{3D}(x,y,z,t)$ the velocity profile, and  finally $S_0 = \tan(a)$ is the absolute value of the flume slope.  We also define the transverse averaged water depth profile:
\begin{equation*}
h(x,t)=\frac{1}{l} \int_{-l/2}^{l/2} h_{2D}(x,y,t) dy,
\end{equation*}
and the transverse and depth averaged velocity profile:
\begin{equation*}
u(x,t)=\frac{1}{l h(x,t)} \int_{-l/2}^{l/2} \int_{0}^{h(x,t)} u_{3D}(x,y,z,t) dy dz.
\end{equation*}
The rain intensity $R(x,t)$ is taken homogenous in space and constant during a duration $t_{stop}$ yielding:
\begin{equation}
R(x,t) = \left\{
\begin{array}{l @{} l}
    & I \ \ \ \text{if}\ t \in [0, t_{stop}] \\
    & 0 \ \ \ \text{if}\ t > t_{stop} \\
\end{array}
\text{for}\ x \in [0, L].
\right.
\label{equ_rain}
\end{equation}

Three dynamical regimes can thus be identified on the measured outflow discharge:
\begin{itemize}
\item between $t = 0\ s$ and a time $t_s$, the water depth in the flume is increasing as well as the outflow discharge: it is the transient, or rising stage,
\item between $t_s$ and $t_{stop}$ the flow is in its steady stage,  and
\item for  $t > t_{stop}$ the rain event is finished and the outflow discharge decreases: it is the recessing stage.
\end{itemize}

This ideal configuration will be studied both experimentally and numerically in order to investigate and validate an effective rainflow overland model.

\subsection{Experimental setup}
\subsubsection{Overall design}
\begin{figure}
\center
\includegraphics[width=0.99\linewidth]{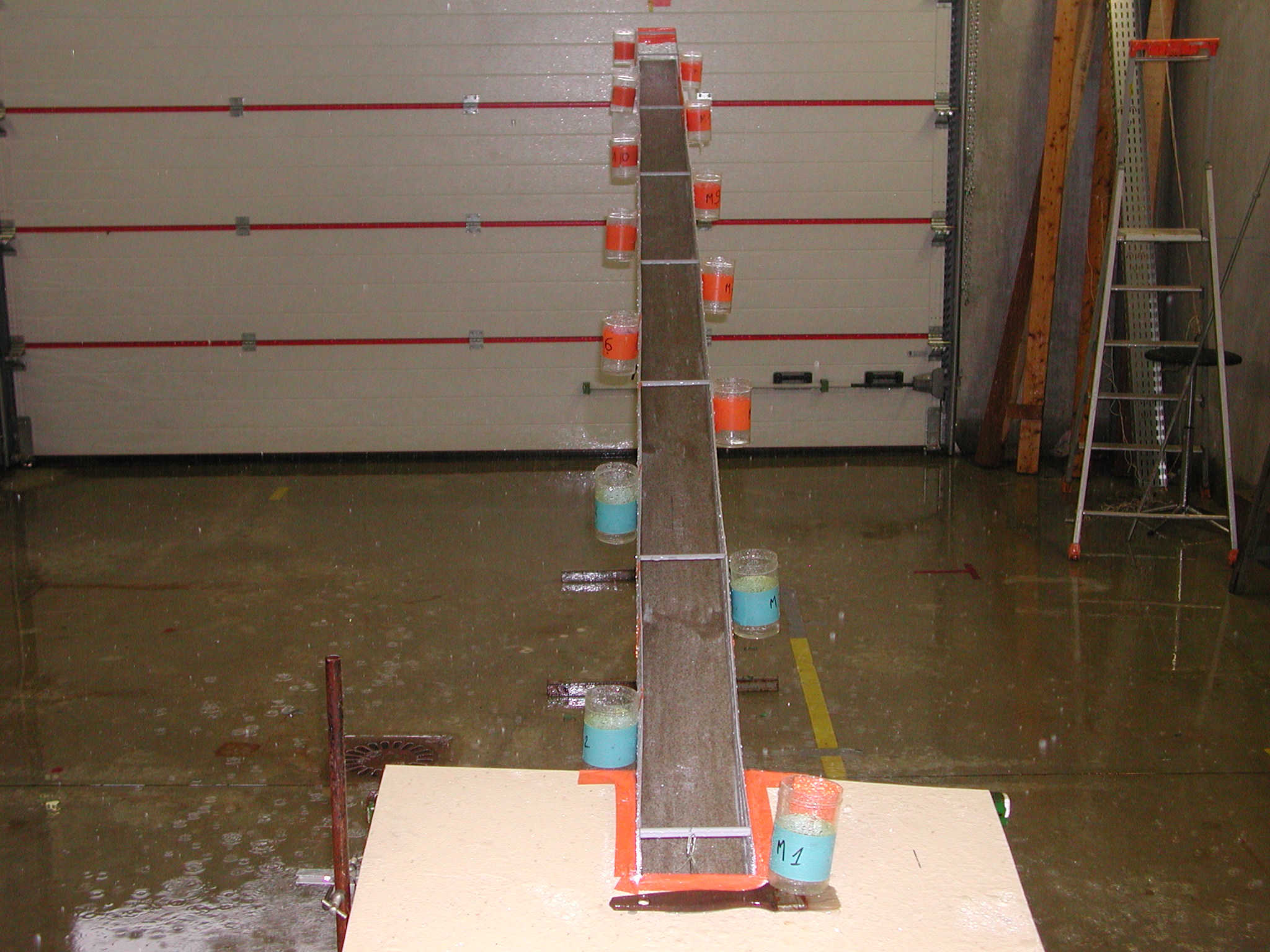}
\caption{Front picture of the flume in the Rainfall Simulation Hall \label{fig_flume}}
\end{figure}

These experiments were carried out at the Rainfall Simulation Hall of the French Institute for Agricultural Research (INRA, Orl\'eans, France). The test bench is a $4.04$ m long and $11.5$ cm wide flat flume having a rectangular section (Fig.~\ref{fig_flume}).  A sheet of glued printing paper is added on the flume for its hydrophilic property, avoiding the formation of threaded flow.  The varying parameters of this experiment are the channel slope $S_0$ and the rainfall intensity. The slope of the panel can be adjusted and is measured using a spirit level (accuracy: $0.5\ mm.m^{-1}$) and a stainless steel rule. The rainfall is produced by a nozzle-type rainfall simulator based on the design of \cite{Foster1979} and located above the channel. Water pressure is set to 90 kPa. Five oscillating nozzles are uniformly distributed over the flume ($1.1\ m$ between them). Using a combination of nozzles with slightly varying openings (Veejet 6540, 6550 and 6560; Spraying System Corp.), a coefficient of variation limited to $8.5 \%$ for the spatial variability of the rain intensity is obtained. Before each experiment, the channel is pre-wetted. A frequency of 55 sweeps per minute is used for the prescribed $50\ mm.h^{-1}$ rainfall intensity (half for the $25\ mm.h^{-1}$).

The experimental cases differences are based on the prescribed rainfall intensity (25 or 50~$\rm mm.h^{-1}$) and slope (2\% or 5\%). 
The three cases considered thereafter are:
\begin{itemize}
\item 25~$\rm mm.h^{-1}$ and 2\%, 
\item 25~$\rm mm.h^{-1}$ and 5\%,
\item 50~$\rm mm.h^{-1}$ and 2\%. 
\end{itemize}
\subsubsection{Measurements}
The data of these measurements can be found in the supplementary material section.
\myparagraph{Outflow hydrograph} 
 The outflow discharge is recorded during the whole run, including both the rising limb of the hydrograph (at the beginning of the rainfall) and its recessing limb (after the end of the rainfall). The outflow discharge is collected in a  bucket by a funnel as schematized in Fig.~\ref{fig_schema}. The outlet of the funnel is custom-made to direct the water flow laterally, avoiding flow pressure to be transmitted to the scale. The cumulative weight of the bucket was recorded using an electronic scale (30 kg range, with a 1 g resolution) at a rate of about 10 Hz. The outflow discharge measurement is replicated six times. The hydrographs (\textit{i.e.} the derivative of the cumulative weight) are quite noisy, because of the high measurement frequency for a small weight increment (maximum flow rate of about $7\ g.s^{-1}$). To make the outflow hydrograph data more readily usable, they are processed by first calculating a moving average over two seconds on each replicate. This duration is long enough to reduce the noise while still being much shorter than the durations of the rising or recessing limbs (which are of several minutes). Then, the median values over the replicates are taken and a Kalman filter (see for instance \cite{Kalman1960}) is applied to smooth the hydrograph.

\myparagraph{Rain intensity} 
During the experimental runs, rainfall intensity is measured by two independent methods:
\begin{itemize}
\item using a set of fourteen beakers positioned along the channel sides and weighted before and after the run,
\item using the flow discharge at steady-state.
\end{itemize}
\myparagraph{Depth and velocity} 
Flow depths and velocities are measured at the middle of the flume width at steady state at up to seven positions along the channel, during one of the replicates. Flow depths are measured using a dial indicator by taking the difference between the reading at the bottom and at the surface. Each flow depth measurement is replicated twice. Flow velocities are measured with the automated salt-tracing gauge described in  \cite{Planchon2005} using  a salt gauge with a 3 cm spacing between the upstream and downstream electrodes. The measurement is carried out for a few minutes at each location, with one reading every ten seconds. 
At each location, for both depth and velocity, the mean value and the standard deviation of the measurements are calculated. This will allow for the comparison between measurements and simulation results.

\subsection{Numerical method}
\subsubsection{Governing equations}
As stated above overland flows are well-described by the Saint-Venant equations, introduced in \cite{DeSaint-Venant1871}, known also  as the non-linear shallow-water equations. These equations are deduced by averaging the Navier-Stokes equations over the water depth, assuming horizontal length scales much larger than the vertical one. In the ``ideal rain" case considered here, the 1D system of Saint-Venant is strictly equivalent of the 2D one because:
\begin{itemize}
\item the topography is constant over the flume width, and 
\item the friction on the walls are not described by the equations.
\end{itemize}
Neglecting the influence of drop impacts on the momentum, the resulting 1D equations of mass and momentum conservation are:
\begin{equation}
\partial_t h(x,t) + \partial_x q(x,t)  = R(x,t) \label{equ_cont} ,
\end{equation}
\begin{equation}
\partial_t q(x,t) + \partial_x \Big( \frac{q(x,t)^2}{h(x,t)} + \frac{g}{2} h(x,t)^2 \Big)  = g h(x,t) ( S_0 - S_f ) ,
\label{equ_mom}
\end{equation}
where $h(x,t)$ and $q(x,t)$ are respectively the local flow depth and the local depth-averaged flux, $R(x,t)$ the rainfall intensity, $g$ the acceleration of gravity, $S_0 =- \partial_x Z_b$ the opposite of the slope (with $Z_b$ the topography) and $S_f$ the friction coefficient in its kinematic form. The derivation of the Saint-Venant equations with rain as the first numerical simulations using this system can be found in \cite{Zhang1989}. We define the maximal Reynolds number  $Re$  with respect to the experimental conditions:
\begin{equation}
Re = \frac{\cos(a) IL}{\nu} ,
\label{eq_reynolds}
\end{equation}
which characterizes the behavior of the fluid : laminar (resp.\ turbulent) for $Re < 500$ (resp.\ $Re > 2000$), where $\nu$ is the kinematic viscosity of the fluid (typically $10^{-6} m^2.s^{-1}$ for water) and $a$ is the angle of the flume with the horizontal.
We define the local Reynolds number with respect to the local value of the numerical 1D fields :
\begin{equation}
Re_l(x,t) = \frac{q(x,t)}{\nu}
\end{equation}
We also introduce the Froude number $Fr$ which characterizes the relative speed of the waves in the flow. The flow is sub-critical (resp.\ supercritical) when the liquid velocity is slower (resp.\ faster) than the surface waves, for $Fr < 1$ (resp.\ $Fr > 1$). The local Froude number is:
\begin{equation}
Fr = \frac{u(x,t)}{\sqrt{g h(x,t)}} .
\label{eq_froude}
\end{equation}
Different friction terms have been proposed in the literature depending on the flow properties. We will consider here the three main friction models: the Darcy-Weisbach model (\textit{e.g.} \cite{Darcy1857}), the Manning model (see for instances \cite{Gauckler1867} and \cite{Manning}), and the Poiseuille model (\textit{e.g.} \cite{Igawaki1955}). The Darcy-Weisbach and Manning models were empirically deduced while the Poiseuille model was obtained analytically.

The Manning model was designed for open channel flows driven by gravity. The friction coefficient follows:
\begin{equation}
S_f^{M}=n^2 \dfrac{q(x,t) \left| q(x,t)\right| }{h(x,t)^{10/3}}  ,
\label{eq_man}
\end{equation}
where $n$ is the Manning coefficient. This coefficient is usually found by a trial and error calibration run.\\
For a laminar flow, the vertical velocity profile is given by a Poiseuille flow. Denoting $u_{2D}(x,z,t)$  the 2D local velocity for a 2D Poiseuille flow and
$$u(x,t)=\frac{1}{h(x,t)} \int_{Z_b}^{h(x,t)} u_{2D}(x,z,t) dz$$
the local depth-averaged horizontal velocity, we can express the 2D local velocity as:
\begin{equation}
u_{2D}(x,z,t)= \frac{3}{2} \frac{u(x,t)}{h^2(x,t)} z (2 h(x,t)-z) .
\end{equation}
A well-known analytical solution of the Poiseuille coefficient $S_f^{P}$, without any free parameter, can be then deduced from the Navier-Stokes equations:  
\begin{equation}
S_f^{P} = \frac{\nu}{g h(x,t)} \partial_z u_{2D}(x,z=0,t) = \dfrac{3 \nu}{g} \frac{q(x,t)}{h^3(x,t)}.
\label{eq_pois}
\end{equation}
Note that in contrast with the Manning models, the Poiseuille friction model does not contain any empirical/adjustable parameter (other than the fluid viscosity which is set to that of water for the case of an ideal rain). 

The Darcy-Weisbach model was initially designed for turbulent flows inside pipes, but it is generally used because the coefficient $f$ can be deduced from the Moody diagram (\textit{e.g.} \cite{Bell1989a}). The friction coefficient for this law can be written in kinematic form as:
\begin{equation}
S_f^{DW}= \dfrac{f}{8 g}\frac{q(x,t) \left| q(x,t)\right| }{h(x,t)^3} ,
\label{equ_darcy}
\end{equation}
where $f$ is the Darcy-Weisbach coefficient. We can find in the literature different laws giving the coefficient $f$ with respect to the local Reynolds number, see for instance the Henderson version (\cite{Henderson1996a}) of the Colebrook-White formulae (\cite{Colebrook1937}), but such laws are not designed to be used for such low Reynolds flows. Here, we propose a simple law for the coefficient $f$ :
\begin{equation}
f = \left\{
\begin{array}{l @{} l}
    & \frac{24}{Re_l} \ \ \ \text{if}\  Re_l < 48 ,\\
    & 0.5 \ \ \ \text{if}\ Re_l \geq 48 . \\
\end{array}
\right.
\label{equ_rain}
\end{equation}
In the low Reynolds region (\textit{i. e.} $Re_l \leq 48$), this law mimics the Poiseuille Model (Equ. \eqref{eq_pois}).
 In the "high" Reynolds region, the value of $f = 0.5$ is chosen to be the highest possible for a smooth surface  (see \cite{Paraschivoiu2003} p.317 for details), in order to have an influence in this setup. 

\subsubsection{Numerical scheme}
Numerical simulations are performed using well-known tested codes that implement the following numerical scheme (\textit{i.e.} \cite{popinet2013} and \cite{Delestre2012a}).
 The shallow-water system of partial derivative equations (PDE) writes under the vectorial form
\begin{equation}
 \partial_t U+ \partial_x F(U)=S(U),
\end{equation}
with
\begin{equation}
 U=\left(\begin{array}{c}
          h(x,t)\\
          q(x,t)
         \end{array}\right),
         \;
F(U)=\left(\begin{array}{c}
            q(x,t)\\
            \frac{gh(x,t)^2}{2}+\frac{q(x,t)^2}{h(x,t)}
            \end{array}\right),
S(U)=\left(\begin{array}{c}
            R\\
            gh(x,t)(S_0-S_f)
            \end{array}\right).            
\end{equation}
This is a set of conservation laws, where the first equation represents the mass conservation and the second one represents the momentum balance. Thus a finite volume method is used which is by construction a conservative method. It consists in integrating
 the equations on cells $[x_{i-1/2},x_{i+1/2}]\times[t^{n},t^{n+1}]$, where $[x_{i-1/2},x_{i+1/2}]$ is centered on point $x_i$.
  We have $x_{i+1/2}-x_{i-1/2}=\Delta x$ and $t^{n+1}-t^n=\Delta t$. After calculations on these cells, with the homogeneous
  system ({\it i.e.} with no rain, no friction and no topography), we get the following explicit in time
   finite volume scheme
     \begin{equation}
     \left\{\begin{array}{l}
     \dfrac{h_i^{n+1}-h_i^n}{\Delta t}+\dfrac{{F_1}_{i+1/2}^n-{F_1}_{i-1/2}^n}{\Delta x}=0\\
     \\
     \dfrac{q_i^{n+1}-q_i^n}{\Delta t}+\dfrac{{F_2}_{i+1/2}^n-{F_2}_{i-1/2}^n}{\Delta x}=0
     \end{array}\right.
     \end{equation}
   where ${F_1}_{i+1/2}^n$ (resp. ${F_2}_{i+1/2}^n$) is the approximation of the first component (resp. the second component)
 of the flux function $F(U)$ at the cells interface located at point $x_{i+1/2}$.
The CFL stability criteria ensure that the scheme is stable for :
\begin{equation}
\Delta t \leq 0.5 \frac{\Delta x}{a} \ \ \text{with} \ \ a =  \max(ap, -am)
\label{eq_CFL}
\end{equation}
where a is the magnitude of the velocity of waves, $ap$ the maximum value of $u_i + \sqrt(G*h_j)$ and $am$ the minimum value of $u_i - \sqrt(G*h_j)$ for $j \in \{i-1;i;i+1\}$ and $ \forall i$ (see \cite{Courant1928} for details).
The topographic term is treated inside the flux term thanks to a well-balanced scheme ({\it i.e.} it captures lake at rest solutions), which is preserving the non-negativity of the water depth ( \cite{Audusse2004a}, \cite{Kurganov2007}). The friction source term is treated semi-implicitly (\cite{Bristeau2001}), the accuracy of the scheme is improved in space with a MUSCL reconstruction (\cite{VanLeer1979}) and in time with a generic second order method (\cite{Williamson1980}).
\subsubsection{Numerical cases}
We simulate a one dimension channel with a fixed slope $S_0 $, as presented in Fig~\ref{fig_schema}. Its horizontal length is  $L_x = \frac{L+2}{\sqrt{1+S_0^2}}$ with $L = 4.04\ m$ and we shift the origin at $X = -1\ m$ to avoid effects of the rain source term at the left boundary. At the right boundary, we put a water tank of 1 meter width and 1 meter depth to reproduce the experimental setup. We set closed boundary condition at the left of the slope ($X = -1\ m$) and at the right ($X = 5.04\ m$). The rain source is equal to zero for $X < 0$ and equal to \eqref{equ_rain} for $X > 0$. 
We chose a reasonably small cell size: $ \Delta x = \frac{L_x}{2096} = 0.00288  \ m$ . The largest time step $\Delta t_{max}$ verifying the CFL condition is automatically chosen by the solver, following the equation \eqref{eq_CFL}. We start the simulation at $t_{start} = 0$ and we stop it at $t_{end} = 1000\ s$. The rain is stopped at $t_{stop} = 600\ s$. 

\begin{figure}
\center
\includegraphics[width=0.99\linewidth]{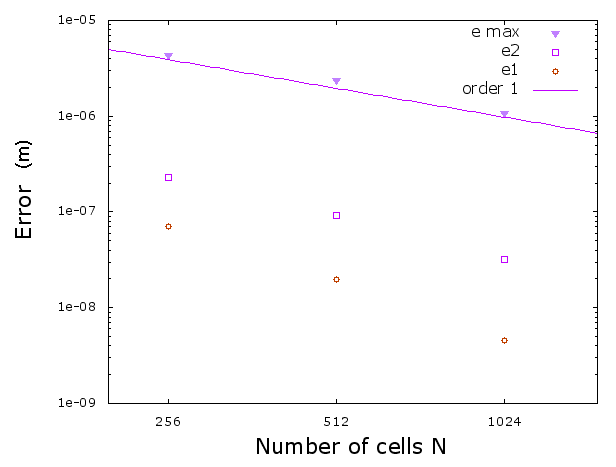}
\caption{Error norms defined in Equ. \ref{equ_e1}, \ref{equ_e2} and \ref{equ_emax} with respect to the number of cells of the simulation calculated for the case ``$I = 25 \rm mm.h^{-1}$ and $S_0 = 5\%$'' for the Darcy-Weisbach friction term. Results shown in log-log scale. The straight line is a guide for the eyes of an order 1 curve.\label{fig_conv}}
\end{figure}

The first stage was to ensure the convergence of simulations.  	
Simulations using the case ``$I = 25 \rm mm.h^{-1}$ and $S_0 = 5\%$'' with different numbers of cells were performed to compute the following error norms at the steady stage (taken at $t = 599\ s$):
\begin{equation}
||e_1(N)|| = \dfrac{\int_0^L |h_N(x) - h_{max}(x)|\ dx}{L},
\label{equ_e1}
\end{equation}
\begin{equation}
||e_2(N)|| = \dfrac{\sqrt{\int_0^L (h_N(x) - h_{max}(x))^2}\ dx}{L},
\label{equ_e2}
\end{equation}
\begin{equation}
||e_{max}(N)|| = max_x (h_N(x) - h_{max}(x)),
\label{equ_emax}
\end{equation}
with $h_N(x)$ the water depth profile with $N$ cells and $h_{max}(x)$ the water depth profile with the maximum number of cells $2096$. We can see in Fig.~\ref{fig_conv}  that our simulations converge. The rate of convergence of $e_{max}$, \textit{i.e.} the maximum error,  is of order one. It is the best convergence rate we can have due to the presence of the shock at the wet-dry transition upstream (Godunov's theorem). 

The second stage prescribes the parameters of the three friction terms. For the Poiseuille friction term, the typical kinematic viscosity $\nu = 10^{-6}\ m^2.s^{-1}$ (water) was considered. As described above, the Poiseuille friction coefficient does not include any calibrated value and the Darcy-Weisbach coefficient depends mainly on the Reynolds number. For the Manning coefficient, a calibration was performed on the experimental case ``$I = 50 \rm mm.h^{-1}$ and $S_0 = 2\%$''. The best possible fit was assessed by trial-and-error. This led to a Manning coefficient of $n = 0.025 \ s.m^{-1/3}$. Thereafter, this value is used for the two other experimental cases. 

\section{Results and discussion}
The parameters relevant to each case are summarized in the Table~\ref{table_results}. For the numerical cases, the rain intensity  ($Num.\ rain$) was chosen to fit the experimental outflow during the steady stage. We also list the values of the Reynolds number and the Froude number computed numerically with the Poiseuille friction term during the steady stage ($t = 599\ s$) at the bottom of the slope ($X=4.04\ m$). Note that the Reynolds number depends only on the experimental conditions. We can see that the flows are always laminar and subcritical. The \textit{``Exp.\ Outflow''} entry in the table is the mean of the discharge measured at the end of the slope during the steady stage for the experimental cases.

\subsection{Hydrographs}
\begin{figure*}
\center
\subfloat[Slope = $2\ \%$, Rain = 25 $mm.h^{-1}$\label{fig_exut2L25}]{%
      \includegraphics[width=0.48\textwidth]{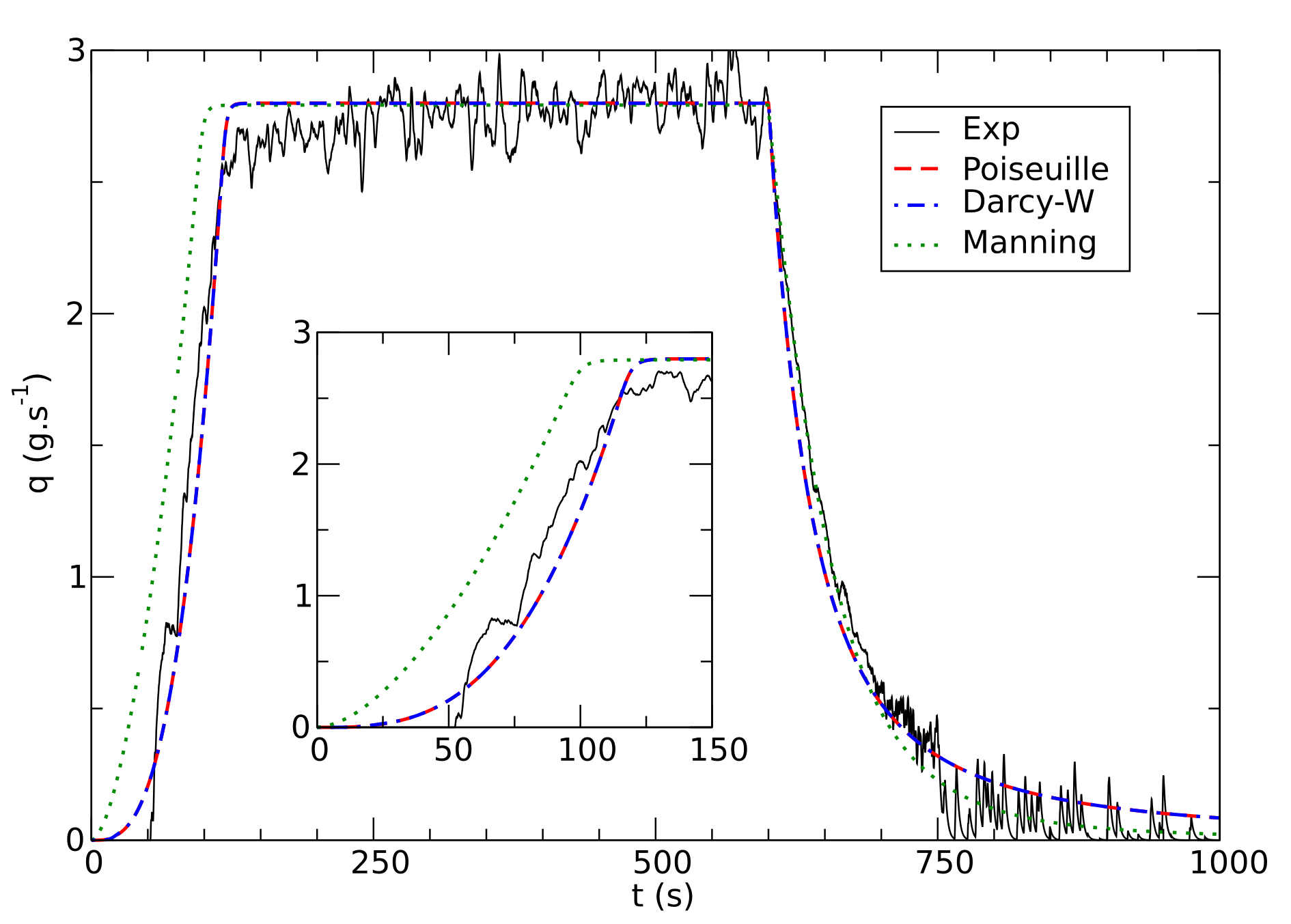}
    }
\hfill
\subfloat[Slope = $5\ \%$, Rain = 25 $mm.h^{-1}$.\label{fig_exut5L25}]{%
      \includegraphics[width=0.48\textwidth]{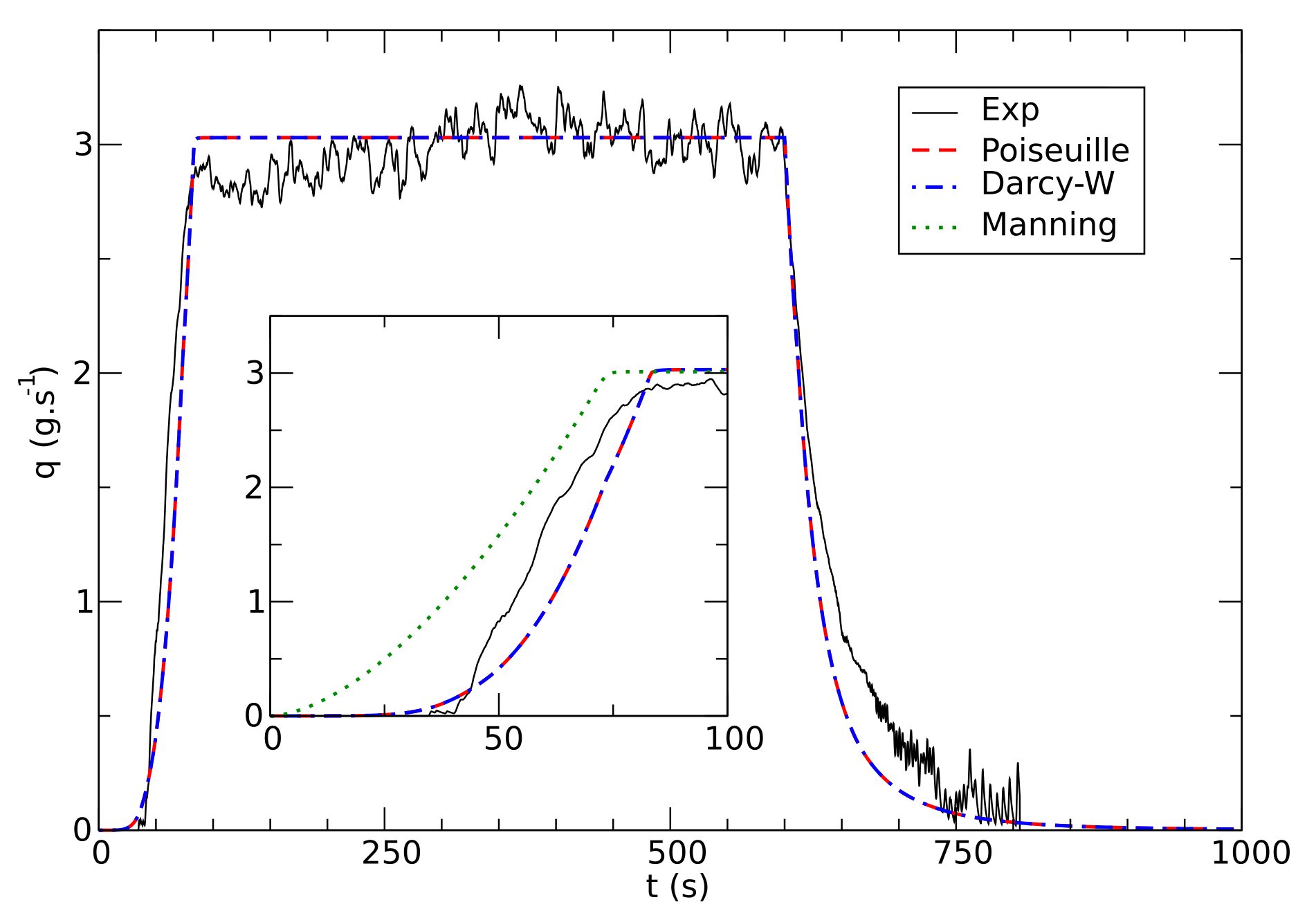}

    }
\vfill
\subfloat[Slope = $2\ \%$, Rain = 50 $mm.h^{-1}$. Definition of $t_b$, $t_s$, $t_{stop}$ and the three stages of the hydrograph. In inset, the Darcy-Weisbach model stops following Poiseuille model at $t=70 \ s$ to follow the Manning model. \label{fig_exut2L50}]{%
      \includegraphics[width=0.8\textwidth]{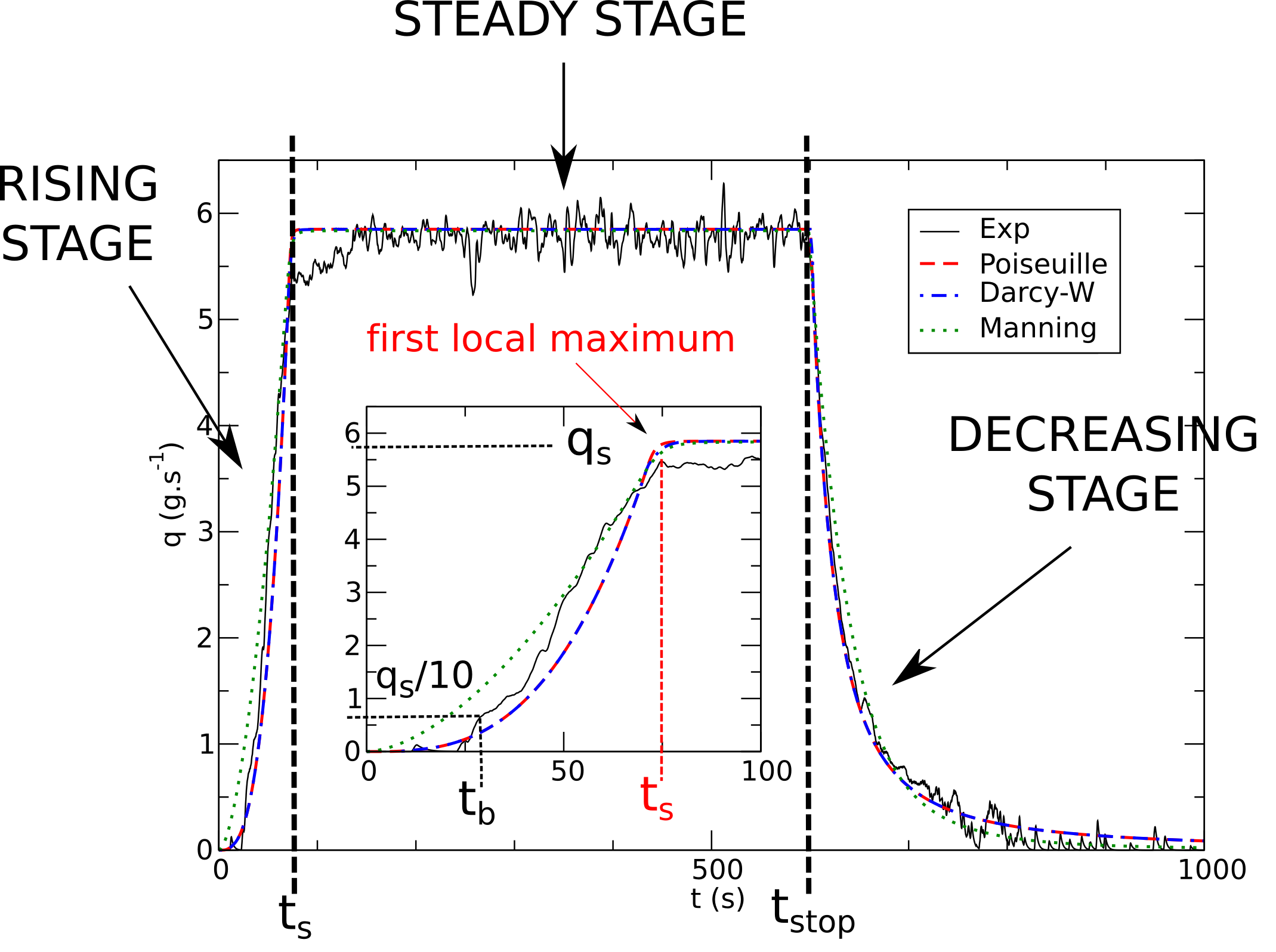}
    }
\caption{Numerical results with different friction terms and experimental discharge at the end of the slope versus time for different slopes and rain intensities. Zoom of the rising limb in inset. \label{fig_exut}}
\end{figure*}

We compute numerically the flow rates at the bottom of the slope for the three different friction terms for  a channel width of 0.115 meter filled with water and we compare them to the experimental measurements. 
The resulting hydrographs for each  case are shown on Fig.~\ref{fig_exut}. 

To illustrate the dynamics of the rising limb, we define two times 
\begin{itemize}
\item $t_b$ as the time when the hydrograph reaches 1/10 of the steady value $q_s$, and
\item $t_s$ as the time when hydrograph reaches its first local maximum, corresponding to the steady state equilibrium.
\end{itemize}
We note on Fig.~\ref{fig_exut5L25} the times $t_b$ and $t_s$  for the experimental case. It is clear that $t_b$ can be considered as the starting time of the rising limb of the hydrograph, and $t_s$ as the beginning of the steady stage. We report on Table~\ref{table_tau} the values of $t_b$ and $t_s$ for each friction term in numerical simulations and for the experimental hydrographs. For the starting time $t_b$, the simulations using the Manning term leads to values much smaller than the experimental value in all cases, while the simulations using the Poiseuille coefficient are much closer. We can see that the simulations using the Darcy-Weisbach model gives similar results than the Poiseuille term, since the local Reynolds number almost never exceeds the critical value (48) of the model. Only for the case $S_0=2\%$ and $I=50mm$ this critical value is reached leading to a small variation only in the results. For the time $t_s$ it is for instance slightly larger than for the Poiseuille model, and no general conclusion can be drawn given such a small effect.  For the beginning of the steady stage $t_s$, the simulations using the Manning term lead to values smaller than expected for the  cases ``$I = 25 \rm mm.h^{-1}$ and $S_0 = 2\%$'' and ``$I = 25 \rm mm.h^{-1}$ and $S_0 = 5\%$'', and to values slightly too high for the case ``$I = 50 \rm mm.h^{-1}$ and $S_0 = 2\%$''. Simulations using the Poiseuille and Darcy-Weisbach friction terms give the closest estimate of $t_s$ for the three experimental cases. Hence, it is clear that the Poiseuille friction term is the best to model the dynamic of the rising stage. Basically, the Manning terms leads to a too early initiation of the rising limb (Fig.~\ref{fig_exut}) while the Darcy-Weisbach term is mimicking  the Poiseuille term in such experiments, except again for the case $S_0 = 2\%$ and $I = 50 mm$  where only a small difference is observed at the end of the rise.
For the steady stage ($ t_s < t <  t_{stop} $), the experimental data shows small oscillations around a mean value because of the water movement in the tank collecting the water flux at the bottom of the slope. The simulated discharges for the three friction terms are strictly equals, because at the steady stage the friction terms do not affect the water flux at the outlet. 

Focusing on the decreasing limb ($t > 600 s$), we observe that, at first, the outflow for Poiseuille decreases faster than for Manning. Then the outflow for Poiseuille becomes higher than for Manning. The Darcy-Weisbach term gives same results as Poiseuille term. However, due to the noise in the experimental hydrographs, it is not really clear which friction term is the best at modeling this stage.

\subsection{Velocity and water depth}

\begin{figure*}
\center
\subfloat[Slope = $2\ \%$, Rain = 25 $mm.h^{-1}$\label{fig_prof2L25}]{%
      \includegraphics[width=0.48\textwidth]{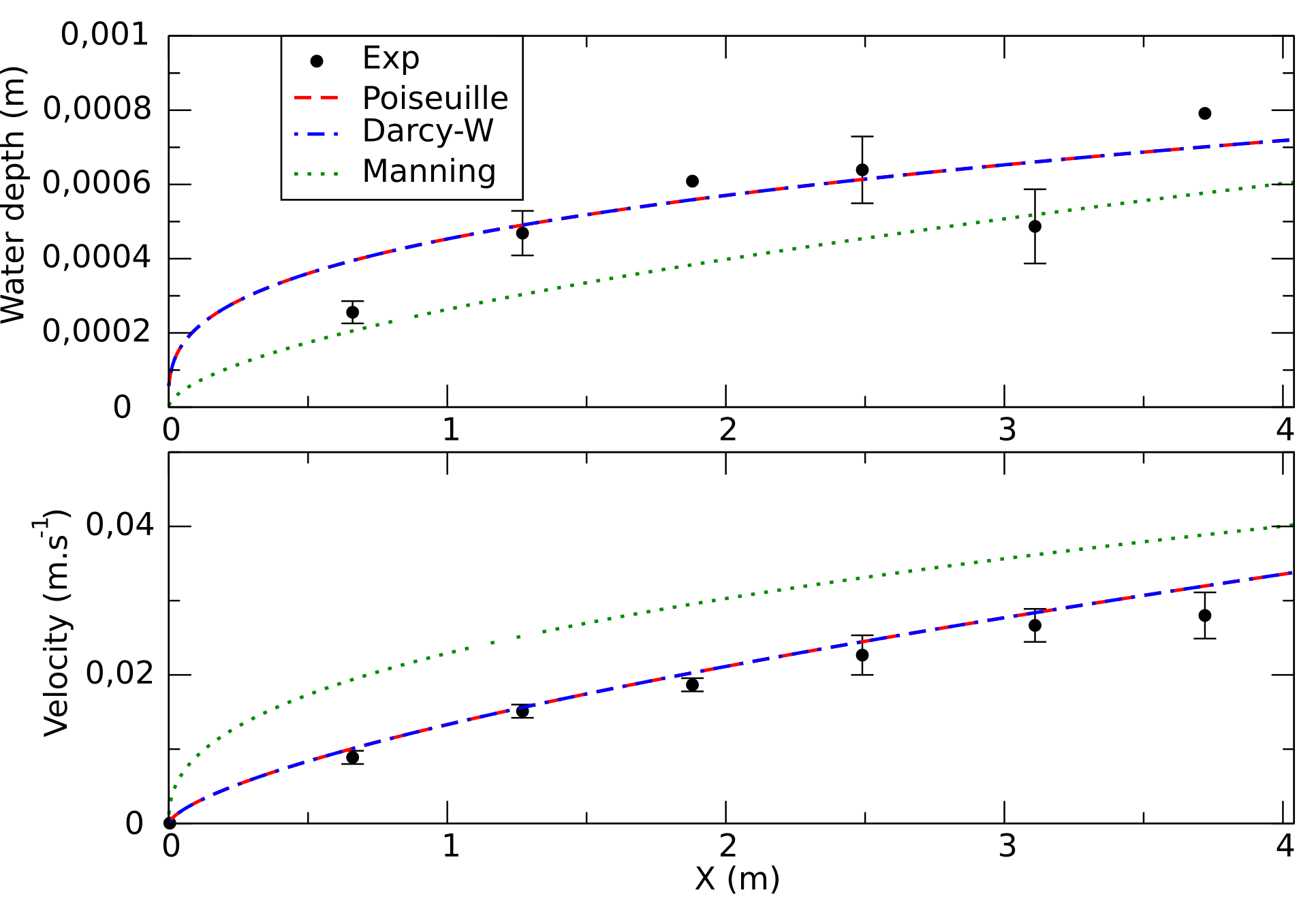}
    }
    \hfill
    \subfloat[Slope = $5\ \%$, Rain = 25 $mm.h^{-1}$\label{fig_prof5L25}]{%
      \includegraphics[width=0.48\textwidth]{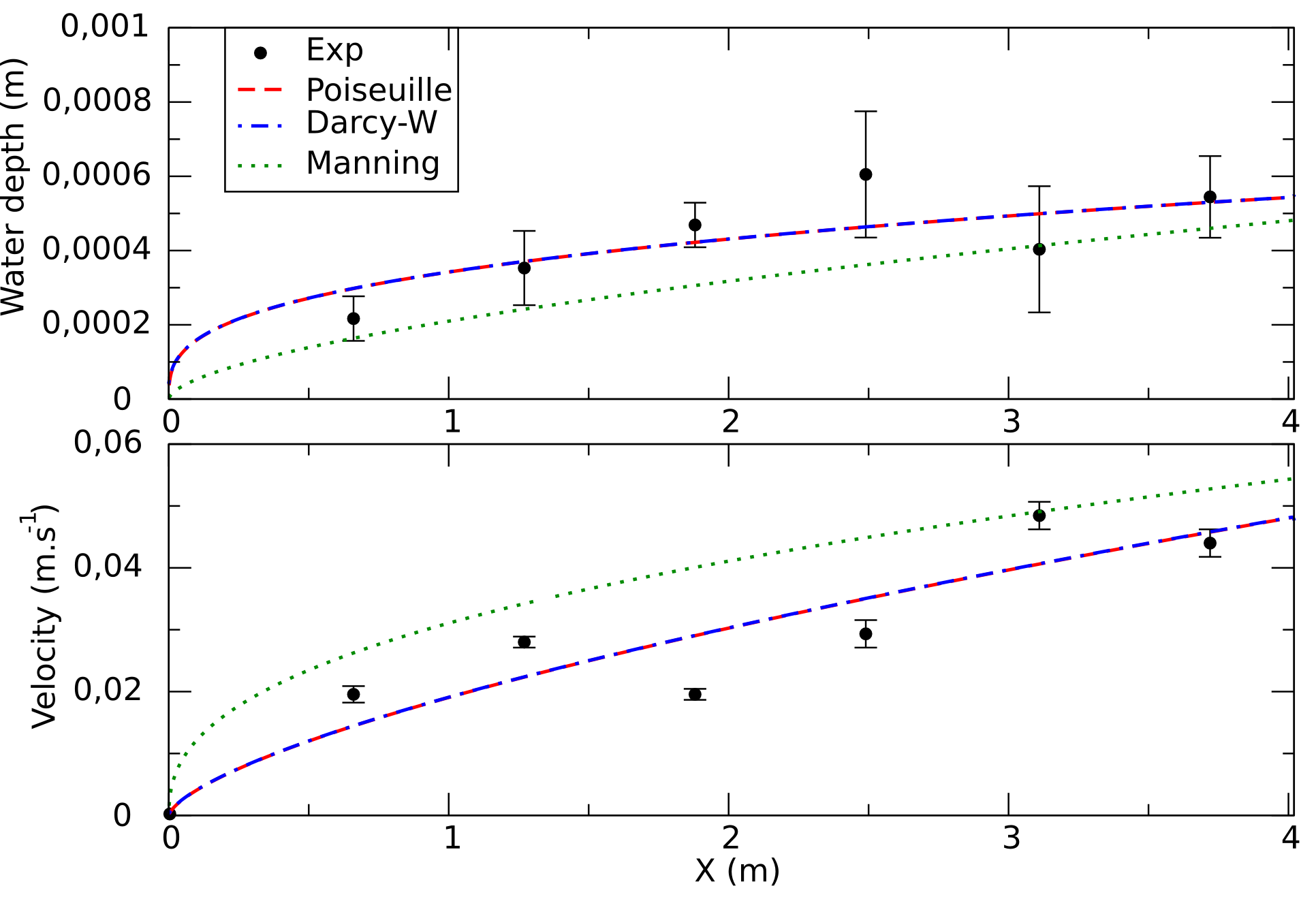}
    }
    \vfill
\subfloat[Slope = $2\ \%$, Rain = 50 $mm.h^{-1}$. At $X = 3.75\ m$, the Darcy-Weisbach model stops following the Poiseuille model and starts following the Manning model.\label{fig_prof2L50}]{%
      \includegraphics[width=0.8\textwidth]{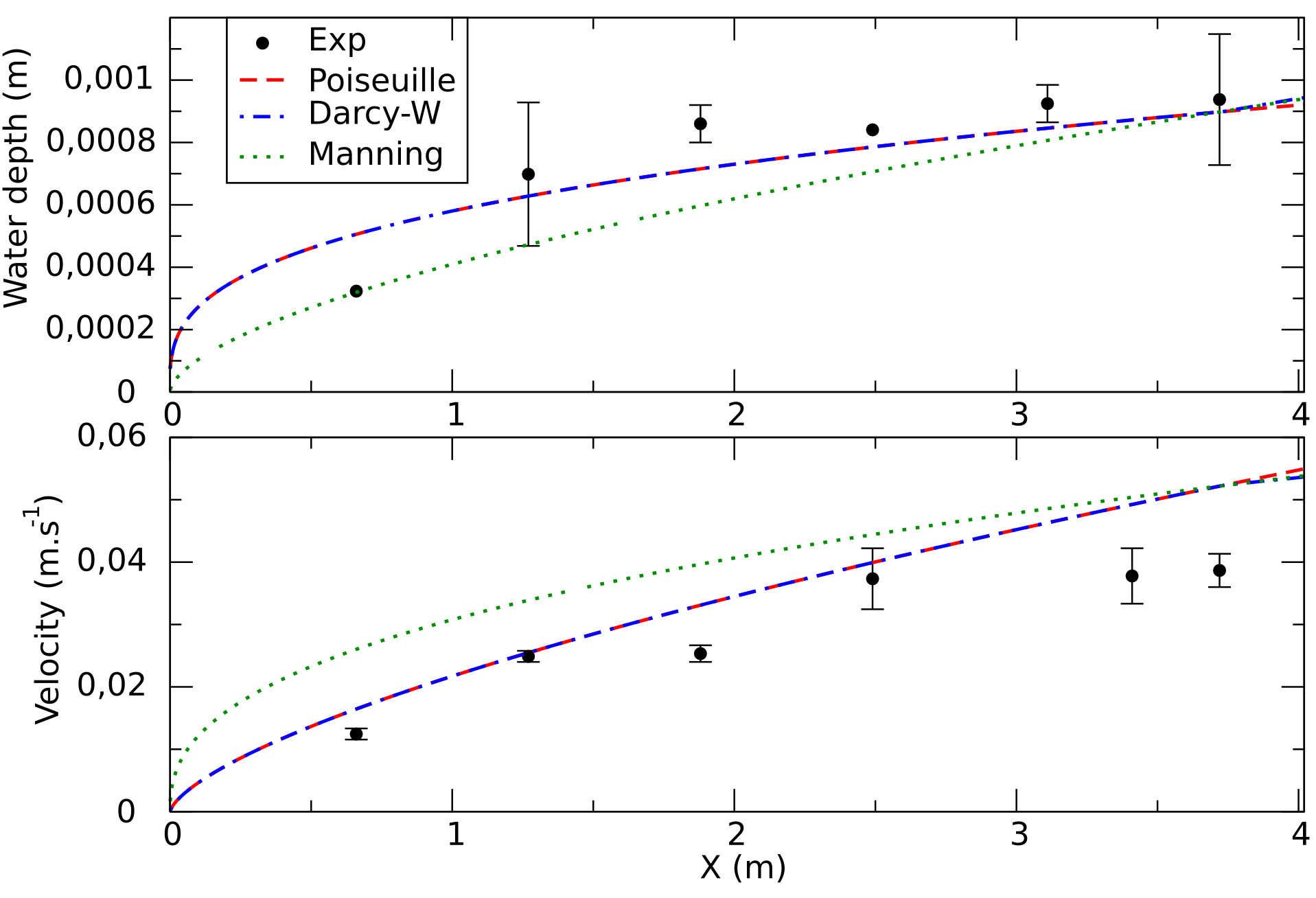}
    }
\caption{Water depth (top) and velocity (bottom) profiles along the slope at the steady stage ($t = 599\ s$).
Error bars are standard errors. \label{fig_prof}}
\end{figure*}

We now look at the velocity profiles for each case during the steady stage ($ t = 599 s$). An important methodological difference is that experimental velocities are measured at the free surface in the middle of the flume, while the 1D numerical profiles can be seen as the transverse averaged values of the $3D$ field. We therefore need to perform some transformations on the velocity field before comparison. Denoting the full $3D$ local velocity field $u_{3D}(x,y,z,t)$,  the 1D velocity profile computed numerically can be expressed $$u(x,t) =\frac{1}{h(x,t) l} \int_{-l/2}^{+l/2} \int_0^{h(x,t)} u_{3D}(x,y,z,t) dy dz.$$
For the 3D velocity profile, we chose as hypothesis a bi-parabolic profile to take into account the influence of walls: 
\begin{equation}
u_{3D}(x,y,z,t) = 9 \frac{u(x,t)}{h^2(x,t) l^2} (\frac{l^2}{4}-y^2) z (2 h(x,t) -z).
\label{equ_vel3D}
\end{equation}
We can finally express the experimental measurement of the velocity with respect to the 1D transverse averaged one as: 
\begin{equation}
u_{3D}(x,y=0,z=h(x,t),t)= \frac{9}{4} u(x,t).
\label{equ_norm_vel}
\end{equation}
We present on Fig.~\ref{fig_prof} the velocity profiles computed numerically and the mean and standard deviation of experimental measurements normalized by $\frac{9}{4}$. Firstly, we can see that the normalized velocity profile is in good agreement with our numerical results independently from the friction law, validating the hypothesis made on the $3D$ velocity profiles in \eqref{equ_vel3D}. However, the Manning velocities are always too large compared to the experimental values. In all three cases, the velocities computed using the Poiseuille term are the closest to the experimental values. To compare the water depth of the numerical simulations against the experimental results, we compute the averaged value of the water depth as: denoting $U_{exp}(X_{bot})$ the closest velocity measurement at the bottom of the slope ($X_{bot} = 3.72\ m$), $\overline{U}_{exp}(X_{bot})$ its transverse averaged value following \eqref{equ_norm_vel} and $h_{exp}(X_{bot})$ the measurement of the water depth at the same coordinates. We compute the flow rates at $X_{bot}$ as: $q_c(X_{bot}) = \overline{U}_{exp}(X_{bot})\times h_{exp}(X_{bot})$.  We can extrapolate the values at the end of the slope $q_c(L)$. During the steady stage, $\partial_t h(x,t) =0$, then solving Equ. \eqref{equ_cont} leads to $q(x) = R\times x$, so that $q_c(L)$ is found using: $q_c(L)=q_c(X_{bot})\times \frac{L}{X_{bot}}$. As already said, we measure the discharge at the end of the slope with the balance and we denote $q_{exp}$ its value during the steady stage. Finally, we normalize the field $h_{exp}$ by a factor: $\frac{q_{exp}}{q_c(L)}$ to find the transverse averaged water depth. With this method, we can extrapolate directly the water depth profile as long as the averaged velocity profile is correct.
For the water depth profiles (Fig.~\ref{fig_prof}), the Manning  term leads to values too low. As for the velocities, the graphics comparison shows that the Poiseuille term gives the best match for all three cases, still with a D-W correction for the case  $S_0 = 2\%$ and $I = 50 mm$. In this case, we can see at $X = 3.75 \ m$ that both water depth and velocity profiles stop to follow the Poiseuille model and start following the Manning model, a trend that is consistent given the experimental results available.  

To make a quantitative assessment of the numerical results, we define for each friction model a water depth index $Ind_h$ and a velocity index $Ind_u$ as follows:
\begin{equation}
Ind_h = \frac{1}{N} \Sigma_{i=1}^N \frac{\sqrt{\left(h_{num}(X_i)-h_{exp}(X_i)\right)^2}}{h_{exp}(X_i)},
\end{equation}
\begin{equation}
Ind_u = \frac{1}{N} \Sigma_{i=1}^N \frac{\sqrt{\left(u_{num}(X_i)-u_{exp}(X_i)\right)^2}}{u_{exp}(X_i)},
\end{equation}
with $N = 6$ the number of experimental measurements, $X_i$ the position on the flume of the experimental measurements, $h_{num}$ and $u_{num}$ the numerical results for the water depth and the velocity, respectively, at the position $X_i$ for the corresponding friction model (Darcy-Weisbach, Manning or Poiseuille) and $h_{exp}$ and $u_{exp}$ the mean of the water depth and velocity, respectively, measured experimentally at the position $X_i$. A zero value for these indexes means that the numerical result fits perfectly the experimental measurements.

Because the experimental measurements are done at left of $X = 3.75\ m$, the Poiseuille and Darcy-Weisbach indices are equals. For the water height, the index is the smallest when the Poiseuille term is used (Table~\ref{table_ind}). Only in the case ``$I = 50 \rm mm.h^{-1}$ and $S_0 = 2\%$'' the Manning term gives a result as good as the Poiseuille term. For the velocity, the index is always the lowest with the Poiseuille term. Hence, it is clear that the Poiseuille friction term is the best to model both the water depth and the velocity profiles at steady state.

Overall, for a smooth surface with a rain-fed, laminar and subcritical flow, the Poiseuille term leads consistently to the best match for the water flux at the outlet during the initiation of the hydrograph, for the water depth profile at steady state and for the velocity profile at steady state.
Hence, the Poiseuille term could be used for inter-rill overland flow, a condition commonly encountered in watershed surface hydrology. The adequacy of this term needs however to be evaluated on field data in the future.

Compared to the empirical Manning term, the Poiseuille term has the advantage to be defined analytically and to have no parameter to be calibrated.  
In watershed surface hydrology, issues of over-calibration, \textit{i.e.} the use of codes requiring the calibration of numerous parameters based on limited data set, have been leading to equifinality cases and to a limited confidence in the simulation quality, as mentioned in \cite{Beven08}. The use of the Poiseuille term could help in achieving a parsimonious parametrization, improving the overall quality of hydrologic simulations.

\section{Conclusion}
Three different friction terms in the Saint-Venant equations have been examined: the commonly used Manning and Darcy-Weisbach models which are empirical and the Poiseuille term, which is deduced directly from the laminar Navier-Stokes equations. The Manning model investigated in this study is using a constant Manning coefficient chosen thanks to a previous trial-and-error run. The Darcy-Weisbach coefficient is following a well-known laminar law at low Reynolds number and a constant value at high Reynolds number, which is set thanks to literature. The Poiseuille term does not depend on any free parameter (aside from the fluid viscosity). The ``ideal rain" case has been reproduced in laboratory and numerical simulations of these events have been performed for these friction terms. The simulation results have been compared with the experimental results. For both the discharge at the end of the flume and for the velocity and water depth profiles along the flume, we have shown that the Poiseuille friction term appears to be the most relevant to reproduce such laboratory experiments. We noted that the Darcy-Weisbach coefficient reproduces the laminar cases investigated here as well as the Poiseuille model. Only small differences are observed for the highest local 
Reynolds situations for which no quantitative conclusions can be drawn. However, such D-W model offers an interesting simple approach able to deal with the variation of the flow structure and should be studied in the future for more turbulent film-flow. On the other hand, the Poiseuille friction term that has been shown to correctly account for laminar film flow needs to be investigated on complex 2D bathymetry for which local slope variations could perturb the laminar approach. Finally, we would like to emphasize that by investigating firstly a simple laminar flow for which both experimental and numerical results could be quantitatively compared, our work paves the road for a systematic approach of complex rain-driven overland flows. 

\section{Acknowledgment}
The Axa Research Fund is thanked for its financial support through a JRI grant. The experimental work was supported by the ANR project METHODE \#ANR-07-BLAN-0232 and was carried out by Lo\"ic Prud'homme and Bernard Renaux, who are thanked for their technical skills.
\section*{References}
\bibliographystyle{elsarticle-harv}

\begin{thebibliography}{37}
\expandafter\ifx\csname natexlab\endcsname\relax\def\natexlab#1{#1}\fi
\expandafter\ifx\csname url\endcsname\relax
  \def\url#1{\texttt{#1}}\fi
\expandafter\ifx\csname urlprefix\endcsname\relax\def\urlprefix{URL }\fi

\bibitem[{An et~al.(2015)An, Yu, Lee, and Kim}]{An2015}
An, H., Yu, S., Lee, G., Kim, Y., 2015. {Analysis of an open source quadtree
  grid shallow water flow solver for flood simulation}. Quaternary
  International, 1--11.

\bibitem[{Audusse and Bristeau(2005)}]{Audusse2004a}
Audusse, E., Bristeau, M.-o., 2005. {A 2d Well-balanced Positivity Preserving
  Second Order Scheme for Shallow Water Flows on Unstructured Meshes}. Journal
  of Computational Physics 206~(1), 311--333.

\bibitem[{Bates et~al.(2010)Bates, Horritt, and Fewtrell}]{Bates2010}
Bates, P.~D., Horritt, M.~S., Fewtrell, T.~J., 2010. {A simple inertial
  formulation of the shallow water equations for efficient two-dimensional
  flood inundation modelling}. Journal of Hydrology 387~(1-2), 33--45.

\bibitem[{Bell et~al.(1989)Bell, Wheater, and Johnston}]{Bell1989a}
Bell, N., Wheater, H., Johnston, P., 1989. {Evaluation of overland flow models
  using laboratory catchment data. II: Parameter identification of physically
  based (kinematic wave) models}. Hydrological sciences journal 34~(3),
  289--317.

\bibitem[{Benkhaldoun et~al.(2007)Benkhaldoun, Elmahi, and
  Sea{\"{\i}}d}]{Benkhaldoun2007}
Benkhaldoun, F., Elmahi, I., Sea{\"{\i}}d, M., 2007. {Well-balanced finite
  volume schemes for pollutant transport by shallow water equations on
  unstructured meshes.} Journal of Computational Physics 226~(1), 180--203.

\bibitem[{Beven(2008)}]{Beven08}
Beven, K., 2008. {On doing better hydrological science}. Hydrological processes
  22~(November 2008), 3549--3553.

\bibitem[{Bristeau and Coussin(2001)}]{Bristeau2001}
Bristeau, M.-O., Coussin, B., 2001. {Boundary Conditions for the Shallow Water
  Equations solved by Kinetic Schemes}. Tech. Rep. RR-4282, INRIA.

\bibitem[{Cai et~al.(2007)Cai, Xie, Feng, and Zhou}]{Cai2007}
Cai, L., Xie, W.~X., Feng, J.~H., Zhou, J., 2007. {Computations of transport of
  pollutant in shallow water}. Applied Mathematical Modelling 31~(3), 490--498.

\bibitem[{Cea et~al.(2010)Cea, Garrido, and Puertas}]{Cea2010}
Cea, L., Garrido, M., Puertas, J., 2010. {Experimental validation of
  two-dimensional depth-averaged models for forecasting rainfall-runoff from
  precipitation data in urban areas}. Journal of Hydrology 382~(1-4), 88--102.

\bibitem[{Chanson(2004)}]{Chanson2004}
Chanson, H., 2004. {The hydraulics of open channel flow : an introduction,
  second edition}, 2nd Edition. Butterworth-Heinemann, Oxford.

\bibitem[{{Chow, V.}(1959)}]{TeChow1959}
{Chow, V.}, T., 1959. {Open channel flow}. MacGraw-Hill Book Co. Inc.,
  New-York.

\bibitem[{Colebrook and White(1937)}]{Colebrook1937}
Colebrook, C.~F., White, C.~M., 1937. {Experiments with Fluid Friction in
  Roughened Pipes}. Proceedings of the Royal Society A: Mathematical, Physical
  and Engineering Sciences 161~(906), 367--381.

\bibitem[{Courant et~al.(1928)Courant, Friedrichs, and Lewy}]{Courant1928}
Courant, R., Friedrichs, K., Lewy, H., 1928. {Uber die partiellen
  Differenzengleichungen der mathematischen Physik}. Mathematische Annalen
  100~(1), 32--74.

\bibitem[{Darcy(1857)}]{Darcy1857}
Darcy, H., 1857. {Recherches exp{\'{e}}rimentales relatives au mouvement de
  l'eau dans les tuyaux (Vol. 1)}. Mallet-Bachelier.

\bibitem[{de~Saint-Venant(1871)}]{DeSaint-Venant1871}
de~Saint-Venant, A.~B., 1871. {Th{\'{e}}orie du mouvement non permanent des
  eaux, avec application aux crues des rivi{\`{e}}res et {\`{a}} l'introduction
  des mar{\'{e}}es dans leurs lit}. Comptes Rendus des s{\'{e}}ances de
  l'Acad{\'{e}}mie des Sciences 73, 237--240.

\bibitem[{Delestre et~al.(2014)Delestre, Cordier, Darboux, Du, James, Laguerre,
  and Planchon}]{Delestre2012a}
Delestre, O., Cordier, S., Darboux, F., Du, M., James, F., Laguerre, C.,
  Planchon, O., 2014. {FullSWOF : a software for overland flow simulation}. In
  Advances in Hydroinformatics, 221--231.

\bibitem[{Delestre et~al.(2009)Delestre, Cordier, James, and
  Darboux}]{Delestre2009}
Delestre, O., Cordier, S., James, F., Darboux, F., 2009. {Simulation of
  Rain-Water Overland-Flow}. In: Proceedings of the 12th international
  conference on Hyperbolic Problems, University of Maryland. College Park
  (USA), pp. 1--11.

\bibitem[{Esteves et~al.(2000)Esteves, Faucher, Galle, and
  Vauclin}]{Esteves2000}
Esteves, M., Faucher, X., Galle, S., Vauclin, M., 2000. {Overland flow and
  infiltration modelling for small plots during unsteady rain: Numerical
  results versus observed values}. Journal of Hydrology 228~(3-4), 265--282.

\bibitem[{Foster et~al.(1979)Foster, Eppert, and Meyer}]{Foster1979}
Foster, G., Eppert, F., Meyer, L., 1979. {A programmable rainfall simulator for
  field plots}. In: Proceedings of Rainfall Simulator Workshop. pp. 45----59.

\bibitem[{Gauckler(1867)}]{Gauckler1867}
Gauckler, P., 1867. {Etudes Th{\'{e}}oriques et Pratiques sur l'Ecoulement et
  le Mouvement des Eaux}. Tech. rep., Gauthier-Villars, Paris.

\bibitem[{Henderson(1996)}]{Henderson1996a}
Henderson, F.~M., 1996. {Open channel flow}. Macmillan, New York.

\bibitem[{Igawaki(1955)}]{Igawaki1955}
Igawaki, Y. K.~U., 1955. {Fundamental studies on the runoff analysis by
  characteristic}. Disaster prevention research institute December~(10), 1--29.

\bibitem[{Kalman(1960)}]{Kalman1960}
Kalman, R.~E., 1960. {A New Approach to Linear Filtering and Prediction
  Problems}. Transactions of the ASME-Journal of Basic Engineering 82~(Series
  D), 35--45.

\bibitem[{Kurganov and Petrova(2007)}]{Kurganov2007}
Kurganov, A., Petrova, G., 2007. {A second-order well-balanced positivity
  preserving central-upwind scheme for the Saint-Venant system}. Communications
  in Mathematical Sciences 5~(1), 133--160.

\bibitem[{Manning et~al.(1890)Manning, Griffith, Pigot, and
  Vernon-Harcourt}]{Manning}
Manning, R., Griffith, J.~P., Pigot, T.~F., Vernon-Harcourt, L.~F., 1890. {On
  the flow of water in open channels and pipes.} Transactions of the
  Institution of Civil Engineers of Ireland 20, 161--207.

\bibitem[{Morgan et~al.(1999)Morgan, Quinton, Smith, Govers, Poesen, Auerswald,
  Chisci, Torri, and Styczen}]{Kinnell1999}
Morgan, R. P.~C., Quinton, J.~N., Smith, R.~E., Govers, G., Poesen, J. W.~A.,
  Auerswald, K., Chisci, G., Torri, D., Styczen, M.~E., 1999. {Discussion on
  'The European soil erosion model (EUROSEM): A dynamic approach for predicting
  sediment transport from fields and small catchments'}. Earth Surface
  Processes and Landforms 24~(6), 563--565.

\bibitem[{Moss et~al.(1979)Moss, Walker, and Hutka}]{Moss1979}
Moss, A.~J., Walker, P.~H., Hutka, J., 1979. raindrop-stimulated transportation
  in shallow-water flows: an experimental study. Sedimentary geology 22,
  165--184.

\bibitem[{Paraschivoiu et~al.(2003)Paraschivoiu, Prud'homme, and
  Robillard}]{Paraschivoiu2003}
Paraschivoiu, I., Prud'homme, M., Robillard, L., 2003. {M{\'{e}}canique des
  fluides}. Presses inter Polytechnique, Paris.

\bibitem[{Planchon et~al.(2005)Planchon, Silvera, Gimenez, {Favis-Mortlock,
  David Wainwright}, {Le Bissonnais}, and Govers}]{Planchon2005}
Planchon, O., Silvera, N., Gimenez, R., {Favis-Mortlock, David Wainwright}, J.,
  {Le Bissonnais}, Y., Govers, G., 2005. {An automated salt-tracing gauge for
  flow-velocity measurement}. Earth Surface Processes and Landforms 30~(7),
  833----844.

\bibitem[{Popinet(2011)}]{Popinet2011}
Popinet, S., 2011. {Quadtree-adaptive tsunami modelling}. Ocean Dynamics
  61~(January), 1261--1285.

\bibitem[{Popinet(2013)}]{popinet2013}
Popinet, S., 2013. http://basilisk.fr.

\bibitem[{Smith(2014)}]{Smith2014}
Smith, M.~W., 2014. {Roughness in the earth sciences}. Earth-Science Reviews
  136, 202--225.

\bibitem[{Smith et~al.(2007)Smith, Cox, and Bracken}]{Smith2007a}
Smith, M.~W., Cox, N.~J., Bracken, L.~J., 2007. {Applying flow resistance
  equations to overland flows}. Progress in Physical Geography 31~(4),
  363--387.

\bibitem[{{Van Leer}(1979)}]{VanLeer1979}
{Van Leer}, B., 1979. {Towards the Ultimate Conservative Difference Scheme}.
  Journal of Computational Physics 32~(1), 101--136.

\bibitem[{Viollet et~al.(1998)Viollet, Chabard, Esposito, and
  Laurence}]{Violet}
Viollet, P.-L., Chabard, J.-P., Esposito, P., Laurence, D., 1998.
  {M{\'{e}}canique des fluides appliqu{\'{e}}e {\'{E}}coulements
  incompressibles}. Presses de l'{\'{E}}cole Nationale des Ponts et
  Chauss{\'{e}}es, Paris.

\bibitem[{Williamson(1980)}]{Williamson1980}
Williamson, J., 1980. {Low-storage Runge-Kutta schemes}. Journal of
  Computational Physics 35, 48--56.
\newline\urlprefix\url{http://www.sciencedirect.com/science/article/pii/0021999180900339}

\bibitem[{Zhang and Cundy(1989)}]{Zhang1989}
Zhang, W., Cundy, T.~W., 1989. {Modeling of Two-Dimensional Overland Flow}.
  Water Resources Research 25~(9), 2019--2035.

\end{thebibliography}

\newpage

\begin{table*}
\center
\begin{tabular}{cccccc}
  \hline
  Tar. Rain  & Slope & Num. Rain & Reynolds & Froude  & Exp. Outflow \\
  ($mm.h^{-1}$) & ($\%$) & ($mm.h^{-1}$) & & & $(g.s^{-1})$  \\ 
  \hline
  25 & 2 & 22 & 24 & 0.4 & 2.8 \\
  \hline
    25 & 5  & 23.5 & 26 & 0.65 & 3.0 \\
    \hline
     50 & 2  & 45.5 & 54 & 0.6 & 5.8 \\
     \hline
 	 \end{tabular}
\caption{\label{table_results} Main quantities for each studied case.}
\end{table*}

\begin{table*}
\center
\begin{tabular}{cccc}
 \hline
  Rain and Slope & Num.\ or Exp.\ Cases & $\ \ t_b\ ($s$)  \ $ & $\ \ t_s\ ($s$) \ $ \\ 
  \hline
  \multirow{4}{*}{25 $mm.h^{-1}$ and 2 $\%$} & Exp. & 55 & 115  \\
    & Poiseuille  & 55 & 120  \\
        & Darcy-W. & 55 & 120 \\
            & Manning  & 30 & 105  \\
    \hline
   \multirow{4}{*}{50 $mm.h^{-1}$ and 2 $\%$} & Exp. & 30 & 75  \\
    & Poiseuille  & 35 & 75  \\
        & Darcy-W. & 35 & 80  \\
            & Manning  & 20 & 80  \\

     \hline
     \multirow{4}{*}{25 $mm.h^{-1}$ and 5 $\%$} & Exp. & 45 & 85  \\
    & Poiseuille  & 40 & 85  \\
        & Darcy-W. & 40 & 85  \\
            & Manning  & 20 & 75  \\
\hline
 	 \end{tabular}
\caption{\label{table_tau} Values of $t_b$ and $t_s$ in each case.}
\end{table*}

\begin{table*}
\center
\begin{tabular}{cccc}
 \hline
  Rain and Slope & \ \ Friction model \ \ & $\ \ Ind_h \ $ & $\ \ Ind_u\ $ \\ 
  \hline
  \multirow{3}{*}{\ 25 $mm.h^{-1}$ and 2 $\%$ \ } &   Poiseuille  &\ \ 0.20 \ \ & \ \ 0.09 \ \ \\
        & Darcy-W. & 0.20 & 0.09 \\
            & Manning  & 0.28 & 0.61  \\
    \hline
   \multirow{3}{*}{50 $mm.h^{-1}$ and 2 $\%$} & Poiseuille  & 0.17 & 0.22  \\
        & Darcy-W. & 0.17 & 0.22  \\
            & Manning  & 0.17 & 0.47  \\
     \hline
     \multirow{3}{*}{25 $mm.h^{-1}$ and 5 $\%$}     & Poiseuille  & 0.17 & 0.23  \\
        & Darcy-W. & 0.17 & 0.23  \\
            & Manning  & 0.25 & 0.37  \\
\hline
 	 \end{tabular}
\caption{\label{table_ind} Values of $Ind_h$ and $Ind_u$ in each case. The closer to zero the index is, the closer to the experimental measurements the simulation is.}
\end{table*}

\end{document}